\documentclass[aip,graphicx,amsmath,amssymb,reprint]{revtex4-1}
\usepackage{graphicx}
\usepackage{dcolumn}
\usepackage{bm}
\usepackage{commath}
\usepackage[utf8]{inputenc}
\usepackage[T1]{fontenc}
\usepackage{mathptmx}
\usepackage{etoolbox}
\usepackage{siunitx}
\DeclareSIUnit{\torr}{Torr}
\DeclareSIUnit{\mTorr}{\milli\torr}
\usepackage{xcolor}

\DeclareUnicodeCharacter{0308}{\"}
\DeclareUnicodeCharacter{030C}{\v{ }}
\DeclareUnicodeCharacter{0301}{\'{ }}

\usepackage{titlesec}
\titleformat{\section}
  {\normalfont\large\bfseries\raggedright}{\thesection.}{0.5em}{}
\titlespacing*{\section}
  {0pt}{3.5ex plus 1ex minus .2ex}{1.5ex plus .2ex}

\titleformat{\subsection}
  {\normalfont\normalsize\bfseries\raggedright}{\thesubsection.}{0.5em}{}
\titlespacing*{\subsection}
  {0pt}{2.5ex plus 1ex minus .2ex}{1ex plus .2ex}
\date{\Today}

\makeatletter
\def\@email#1#2{%
 \endgroup
 \patchcmd{\titleblock@produce}
  {\frontmatter@RRAPformat}
  {\frontmatter@RRAPformat{\produce@RRAP{*#1\href{mailto:#2}{#2}}}\frontmatter@RRAPformat}
  {}{}
}
\makeatother
\draft

\begin{document}

\title{Near-zero effective magnetization enabling ultra-low threshold currents in spin Hall micro-oscillators}

\author{A. Koujok*}
\affiliation{Fachbereich Physik and Landesforschungszentrum OPTIMAS, Rheinland-Pf\"alzische Technische Universit\"at Kaiserslautern-Landau, Kaiserslautern, Germany}
\email{a.koujok@rptu.de}
\author{H. Kurebayashi} \affiliation{London Centre for Nanotechnology, University College London, London, United Kingdom}
\affiliation{Department of Electronic and Electrical Engineering, University College London, London, United Kingdom}
\affiliation{WPI Advanced Institute for Materials Research, Tohoku University, Sendai, Japan}
\author{K. Yamamoto} \affiliation{Advanced Science Research Center, Japan Atomic Energy Agency, Tokai, Japan}
\author{B. Heinz} \affiliation{Fachbereich Physik and Landesforschungszentrum OPTIMAS, Rheinland-Pf\"alzische Technische Universit\"at Kaiserslautern-Landau, Kaiserslautern, Germany}
\author{V. K. Kushwaha} \affiliation{Institute for Materials Research, Tohoku University, Sendai, Japan}
\author{X. Hou} \affiliation{Institute for Materials Research, Tohoku University, Sendai, Japan}
\author{A. Hamadeh}\affiliation{Centre de Nanosciences et de Nanotechnologies, CNRS, Universit\'e Paris-Saclay, Palaiseau, France}
\author{T. Seki} \affiliation{Institute for Materials Research, Tohoku University, Sendai, Japan} \affiliation{Center for Science and Innovation in Spintronics, Tohoku University, Sendai, Japan} \affiliation{International Center for Synchrotron Radiation Innovation Smart, Tohoku University, Sendai, Japan}
\author{P. Pirro} \affiliation{Fachbereich Physik and Landesforschungszentrum OPTIMAS, Rheinland-Pf\"alzische Technische Universit\"at Kaiserslautern-Landau, Kaiserslautern, Germany}

\date{\today}

\begin{abstract}

Reducing the electrical current required to excite magnetization dynamics is a central challenge, e.g. for energy-efficient magnonic devices or oscillator-based computing. Spin Hall oscillators typically rely on large current densities to compensate intrinsic magnetic damping, so these systems are usually studied on the nanoscale (Spin Hall Nano Oscillators, SHNOs) to work with moderate currents and a favorable heat dissipation geometry. Here, we demonstrate that engineering a near-zero effective magnetization ($M_\mathrm{eff}$) enables a drastic reduction of the magnetization oscillation threshold current density for Spin Hall oscillators. This makes it possible to excite even comparably large systems with micrometer lateral sizes, so-called "Spin Hall Micro-Oscillators" (SHMOs). Using micro-focused Brillouin light scattering spectroscopy, we quantify the threshold current density in SHMOs based on W/CoFeB/MgO/Ta  with near-zero $M_\mathrm{eff}$. We observe threshold current densities as low as $J_{\mathrm{th}}$ = (0.292 $\pm$ 0.025) $\times 10^{10}$ \,\si{\ampere\per\square\metre}, representing a reduction of more than two orders of magnitude compared with most recent reported SHNOs. Using systematic micromagnetic simulations, we investigate the breaking down of the macrospin approximation and underline the high influence of $M_\mathrm{eff}$ on magnetization dynamics under applied spin currents. Our results establish $M_\mathrm{eff}$ engineering as a powerful strategy for realizing ultra-low-power spin Hall oscillators and energy-efficient magnetization control.

\end{abstract}


\maketitle

The rising need for unconventional computation means drives research in magnonics, a field where spin-waves and their quanta magnons are investigated for their intrinsic non-linearity and tunability \cite{chumak2022advances,fischer2017experimental,csaba2017perspectives,mahmoud2023two,levchenko2024review,kohl2025identification}. Spin Hall oscillators \cite{demidov2014synchronization,awad2017long,li2020recent,choi2022voltage} are non-linear micro/nano-sized magnonic devices offering promising applications in microwave generation\cite{liu2013spectral,chen2016spin} and unconventional computation \cite{zahedinejad2020two,zahedinejad2022memristive}. These devices typically consist of a bi-layer whereby a heavy metal is in contact with a magnetic material. Due to spin-orbit coupling in the heavy metal and inversion symmetry breaking due to the interface, charge currents flowing in the heavy metal give rise to a pure spin current via the spin Hall effect (SHE) \cite{demidov2012magnetic,edwards2012parametric,yasuda2017current,li2020recent}. That pure spin current flows transverse to the direction of the flow of electrons and into the adjacent magnetic material causing a destabilization of its magnetization by spin orbit torques which can lead to sustained oscillations or magnetization reversal \cite{demidov2012magnetic,awad2017long}. While promising, spin Hall oscillators face challenges in terms of energy consumption, as high current densities are usually needed to drive magnetization dynamics and generate microwave signals \cite{haidar2019single}, thus resulting in heating due to increased Ohmic losses. The process of overcoming these challenges employs sophisticated fabrication techniques and material engineering, as the efficiency of spin Hall systems in terms of energy conversion heavily depends on how the stacks are fabricated and on selecting best fit materials that constitute those stacks. Indeed, studies have been and are still being conducted to improve the efficiency of these devices, for instance by engineering means of increasing the spin Hall angle \cite{jungwirth2012spin,haidar2021compositional}, by miniaturizing their sizes into nano-constrictions \cite{behera2024ultra} or through using special materials \cite{bainsla2025energy}.\\
In this work, we utilize a spin Hall oscillator with near zero effective magnetization ($M_{\mathrm{eff}}$) to experimentally realize extremely low oscillation threshold current densities $J_\mathrm{th}$. This makes it possible to excite even comparably large systems with micrometer lateral sizes, systems we refer to as "Spin Hall Micro-Oscillators" (SHMOs), using low currents. We employ micro-focused Brillouin light scattering spectroscopy ($\mu$BLS) to quantify $J_\mathrm{th}$, the current density required for the onset of the driven oscillations. Compared to other recent advances in this direction \cite{behera2024ultra,bainsla2025energy}, the value of $J_\mathrm{th}$ found in our measurements exhibits a reduction of more than two orders of magnitude. We attribute this reduction to the strongly reduced precessional ellipticity at low $M_\mathrm{eff}$ \cite{divinskiy2019controlled,guckelhorn2021magnon}, which leads to low magnon Gilbert damping rates and also suppresses damping induced by magnon scattering processes due to a reduced magnon-magnon interaction coefficient. Additionally, we study via micromagnetic simulations the process of breaking down of the macrospin model in terms of describing magnetization dynamics in SHMOs, and highlight the effect of $M_\mathrm{eff}$ on magnetization dynamics beyond macrospin. This paves way to consider these systems as a promising platform to demonstrate novel effects like the magnonic Klein paradox \cite{harms2022enhanced,yuan2023magnonic} in addition to further effects based on antimagnons and dynamical stability \cite{harms2022enhanced,harms2024antimagnonics,kurebayashi2026dynamical,karadza2026dynamical,Wang2026direct}.

\begin{figure}[!ht]
\centering
\includegraphics[width=\columnwidth]{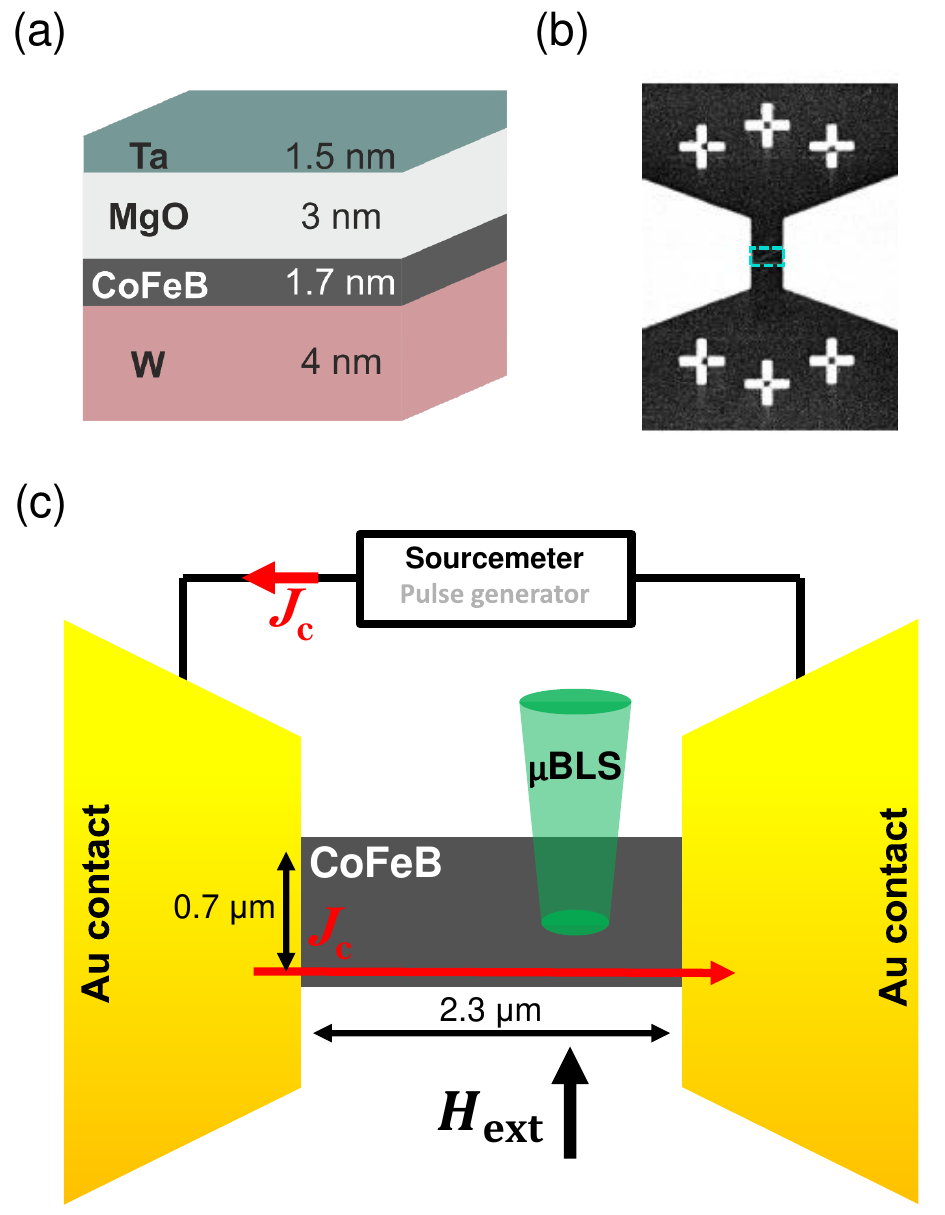}
\caption{(a) Schematic representation of the sample stack. A W layer of thickness 4 \,\si{\nano\metre} is used for SHE. The latter is in contact with an ultra thin CoFeB layer with a thickness of 1.7 \,\si{\nano\metre}. The CoFeB is capped with 3 \,\si{\nano\metre} of MgO. The MgO layer is topped with a Ta layer with a thickness of 1.5 \,\si{\nano\metre}. (b) SEM image displaying a lateral view of the SHMO (cyan highlighted region), the Au contact pads connected to it and the plus-shaped markers used to stabilize the micro-focusing at positions of interest. (c) Simplified schematic of the setup used to carry-out the experimental measurements. The micro-focusing of the laser is shown as a green funnel. A sourcemeter/pulse generator is used to create the potential difference at the Au contacts. $J_{\mathrm{c}}$ denotes the charge current supplied by the sourcemeter/pulse generator. The magnetic field $H_{\mathrm{ext}}$ is applied perpendicular to the flow of electric charges. Magnon intensity and frequency can then be measured after the inelasticlly scattered light is gathered.}\label{fig1}
\end{figure}

In FIG. \ref{fig1} (a), we present the material stack whereby the ferromagnetic layer composed of 1.7 \,\si{\nano\metre} thick Cobalt-Iron-Boron (CoFeB) is sandwiched between the heavy metal used to generate the SHE, namely the Tungsten (W) layer at one end, and a Magnesium Oxide (MgO) layer followed by a Tantalum (Ta) capping at the other end. Tungsten is a typical heavy metal used in such devices for its relatively large spin Hall angle in comparison to other heavy metals \cite{hao2015beta,behera2022energy}. The interface between MgO and CoFeB creates a perpendicular magnetic anisotropy (PMA) along the out-of-plane direction. Since the effective magnetization is defined as $M_\mathrm{eff}=$ $M_\mathrm{s}$ - $\frac{2 K_u}{\mu_0 M_\mathrm{s}}$, $K_\mathrm{u}$ being the uniaxial out-of-plane anisotropy constant, the result is hence a near-fully compensated $M_{\mathrm{eff}}$. For more information on the fabrication and sample processing procedures, refer to the \textbf{Stack Fabrication} subsection of the \textbf{Methods} section. The aforementioned compensation of $M_{\mathrm{eff}}$ reduces non-linear effects such as magnon scattering events \cite{divinskiy2019controlled,guckelhorn2021magnon}, and should hence facilitate initiating and controlling magnetization dynamics at low applied currents, and in other words reduce the threshold for magnetization oscillations. The decreased magnon-magnon interaction makes it seem sensible to consider a macrospin model for systems driven by spin torque \cite{slavin2009nonlinear}, where the threshold current density $J_{\mathrm{th}}$ for compensating the system's Gilbert damping and initiating oscillations depends on the magnetic parameters as follows:

\begin{align}
    J_{\mathrm{th}} = 
\frac{2 e \mu_{\mathrm{0}} \alpha d_{\mathrm{FM}} M_s}{\hbar \, \theta_{\mathrm{SH}}}
\left( H_{\mathrm{ext}} + \frac{M_{\mathrm{eff}}}{2} \right)
\label{eq:1}
\end{align}

\noindent where $e$ is the elementary electron charge, $\mu_{\mathrm{0}}$ is the permeability of vacuum, $\alpha$ is the Gilbert damping constant, $d_\mathrm{FM}$ is the thickness of the ferromagnetic material, $M_\mathrm{s}$ is the saturation magnetization, $\hbar$ is the reduced Planck's constant, $\theta_\mathrm{SH}$ is the spin Hall angle, $H_\mathrm{ext}$ is the external magnetic field and $M_\mathrm{eff}$ is the effective magnetization \cite{taniguchi2016instability,zhu2020threshold}. To optimize energy conversion in spin Hall oscillators, and to reduce threshold current densities various approaches have indeed been explored. The effect of the composition and thickness of the heavy metal layer on the threshold current has been previously investigated \cite{liu2012spin,behera2022energy}. The increase in effective damping due to adding Platinum layers (heavy metal) and its effect on increasing the threshold has been demonstrated in \cite{haidar2016controlling}. Lowering the damping compensation threshold by using magnetic insulators with low intrinsic damping has been investigated in \cite{collet2016generation}. Voltage control of magnetic anisotropy and resultant effects on the effective field and the effective magnetization have been studied in \cite{fulara2020giant}. It is evident from Eq. (\ref{eq:1}) that lowering $M_\mathrm{eff}$ will lower $J_{\mathrm{th}}$, and the effect of low $M_\mathrm{eff}$ on faster magnetization switching has been indeed investigated both experimentally and theoretically before \cite{shi2018fast,zhu2020threshold}. Here, we experimentally investigate and quantify the magnetization magnetization oscillation threshold reduction as a result of near zero $M_\mathrm{eff}$.

In FIG. \ref{fig1} (b), a scanning electron microscopy (SEM) image is presented. The SHMO is highlighted in the cyan area (lateral view) between the large Gold contacts used to inject the current. The plus-shaped markers are used stabilize the micro-focusing such that auto-focusing process preserves a fixed position. Figure \ref{fig1} (c) is a simplified schematic of the experimental setup and the studied system. Shown in gray is the CoFeB micro-bar from a top view. The latter has a length of 2.3 \,\si{\micro\metre} and a width of 0.7 \,\si{\micro\metre}. A source-meter or a pulse generator is used to create a potential difference at the Gold (Au) contacts. This results in a continuous or short-pulsed charge current flow ($J_{\mathrm{c}}$), respectively, as shown in the red arrows. The resulting spin current is then injected into the CoFeB. $H_{\mathrm{ext}}$ is applied along the CoFeB micro-bar's short axis. A green laser with a wavelength of 532 \,\si{\nano\metre} is used to optically investigate the resulting magnetization dynamics. The green funnel highlights the illuminated spot by the micro-focused laser.\\

\section*{Results}

\begin{figure}[!ht]
\centering
\includegraphics[width=0.9\columnwidth]{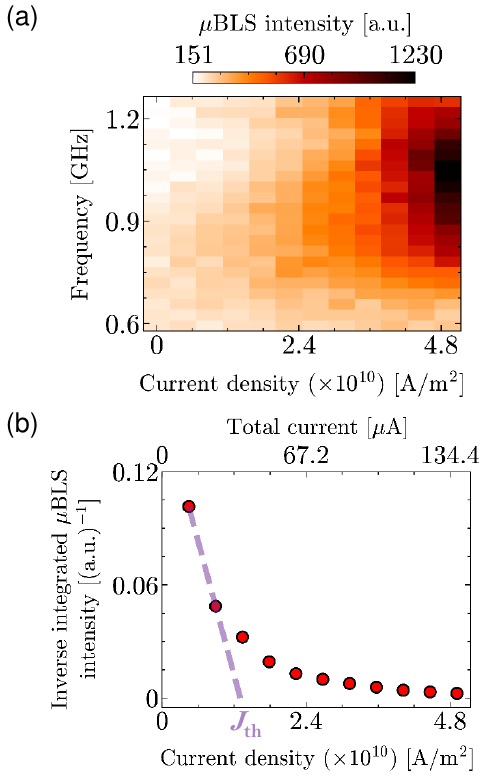}
\caption{Continuous DC measurements: (a) color-coded $\mu$BLS intensity as a function of the applied current density $J_{\mathrm{c}}$ and the $\mu$BLS frequency. (b) The inverse $\mu$BLS intensity as a function of $J_{\mathrm{c}}$ highlighting the threshold current density ($J_{\mathrm{th}}$) in purple, with $J_{\mathrm{th}}$ = (1.305 $\pm$ 0.153) $\times 10^{10}$ \,\si{\ampere\per\square\metre}. All measurements are performed at $\mu_{\mathrm{0}}H_{\mathrm{ext}}$ = 25 \,\si{\milli\tesla}.}\label{fig2}
\end{figure}

\textbf{Continuous DC current measurements.} First, continuous DC measurements are carried out in which a source-meter is used to supply the current. The injected continuous DC current density is swept from 0 up to 4.8 $\times 10^{10}$ \,\si{\ampere\per\square\metre} at a fixed field $\mu_{\mathrm{0}}H_{\mathrm{ext}}$ = 25 \,\si{\milli\tesla}, whilst focusing the laser at the same spot on the CoFeB micro-bar. Figure \ref{fig2} (a) is a 2D map of the $\mu$BLS intensity as a function of $\mu$BLS frequency and current density ($J_{\mathrm{c}}$). For the range of densities below 2.4 $\times 10^{10}$ \,\si{\ampere\per\square\metre}, the detected µBLS intensity remains close to the thermal background. Increasing $J_{\mathrm{c}}$ from 2.4 $\times 10^{10}$ \,\si{\ampere\per\square\metre} to 4.8 $\times 10^{10}$ \,\si{\ampere\per\square\metre}, the signal gradually intensifies meaning that the current is becoming high enough to contribute to a larger enhancement in the thermal magnon signal below the threshold, and to a compensation of the intrinsic damping at and above the threshold. It can be seen that the measured spin-wave spectrum is quite broad and covers nearly a 0.5 \,\si{\giga\hertz} range of frequencies. At large enough current densities, near 4.8 $\times 10^{10}$ \,\si{\ampere\per\square\metre}, the measured spectrum appears to have a well defined maximum around 1 \,\si{\giga\hertz}. We attribute the spectral broadening partially to residual inhomogeneities of the material parameters (e.g. of the out-of-plane anisotropy constant) and the spatial averaging over the  multiple spin-wave modes confined within the 300 \,\si{\nano\metre} laser spot due to laser heating effects \cite{birt2013brillouin}.

Next, the method of extracting the threshold current density $J_{\mathrm{th}}$ from this DC measurement is shown in FIG. \ref{fig2} (b). Below the threshold, the average energy stored in the i-th spin-wave mode ($\langle E_{\mathrm{i}} \rangle$) is directly proportional to the $\mu$BLS intensity $I_\mathrm{BLS}$  \cite{demidov2011control} :

\begin{align}
    I_\mathrm{BLS}\propto\langle E_{\mathrm{i}} \rangle \propto k_{\mathrm{B}} T 
\frac{\Gamma_{\mathrm{i}}}{\Gamma_{\mathrm{i}} - \Gamma({J_\mathrm{c}})}
\label{eq:2}
\end{align}

\noindent with $k_{\mathrm{B}}$ being the Boltzmann constant, $T$ the Temperature, $\Gamma_{\mathrm{i}}$ the relaxation rate of the spin-wave mode and $\Gamma({J_\mathrm{c}})$ the growth rate due to the influence of the spin-orbit torque which depends on the applied current density. The net relaxation rate $\Gamma_{\text{net}}$ is then defined as $\Gamma_{\text{net}}$ $=$ $\Gamma_{\mathrm{i}}$ $-$ $\Gamma({J_\mathrm{c}})$, with the threshold current density being the point of zero net relaxation rate. Considering (\ref{eq:2}), the inverse $\mu$BLS intensity then tends to zero as the net relaxation rate approaches zero. Figure \ref{fig2} (b) depicts this relation by showing the decrease of the inverse intensity as a function of the increasing $J_{\mathrm{c}}$. A linear interpolation of the data points (purple dotted line) at low currents (below the oscillation threshold) allows to quantify an oscillation threshold of $J_{\mathrm{th}}$ = (1.305 $\pm$ 0.153) $\times 10^{10}$ \,\si{\ampere\per\square\metre}. This value represents a reduction of more than an order of magnitude in terms of current density needed to initiate and reach non-decaying oscillations when compared to recent published results in comparable conditions \cite{behera2024ultra,bainsla2025energy}. From a physical perspective, this reduction can be attributed to the reduction of magnon-magnon scattering resultant of the strongly reduced precessional ellipticity at low $M_\mathrm{eff}$ \cite{divinskiy2019controlled,guckelhorn2021magnon}.

\begin{figure*}[ht]
\centering
\includegraphics[width=\textwidth]{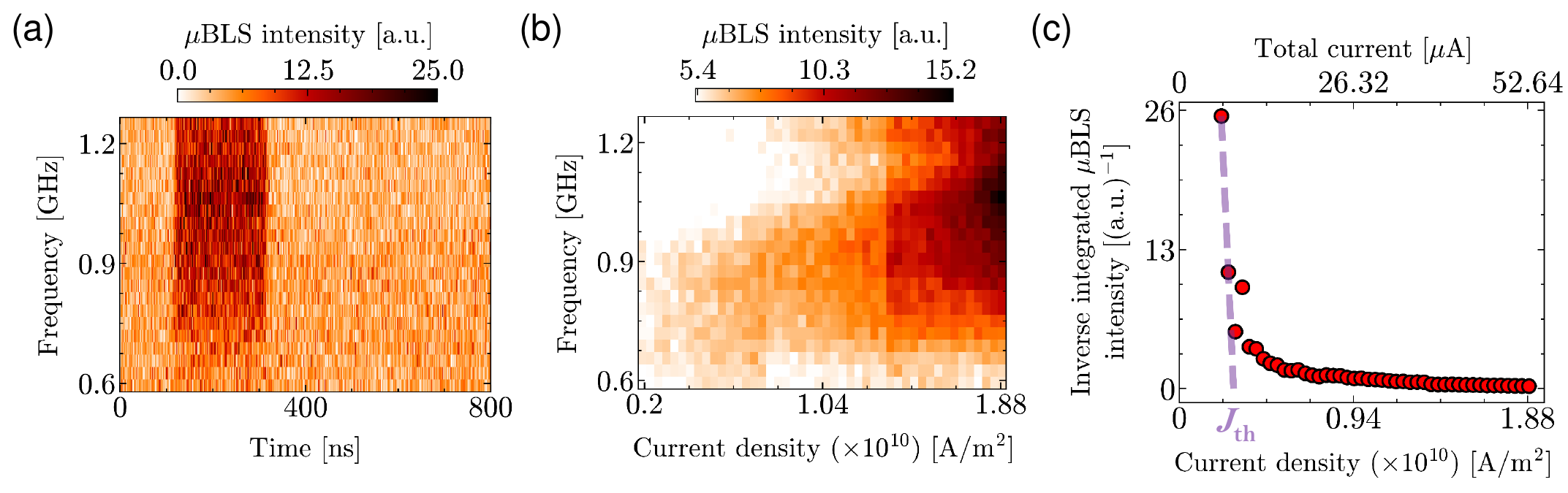}
\caption{Pulsed current measurements: (a) Color-coded $\mu$BLS intensity map of the time resolved BLS frequency. The pulse 200 \,\si{\nano\second} long pulse is highlighted as the increase in intensity in the interval ranging from 100 \,\si{\nano\second} to 300 \,\si{\nano\second}. (b) Color-coded $\mu$BLS intensity map as function of BLS frequency and the applied current density during the pulse. (c) Inverse $\mu$BLS intensity versus current density plot highlighting $J_{\mathrm{th}}$, with $J_{\mathrm{th}}$ = (0.292 $\pm$ 0.035) $\times 10^{10}$ \,\si{\ampere\per\square\metre}. All measurements were carried-out at $\mu_{\mathrm{0}}H_{\mathrm{ext}}$ = 25 \,\si{\milli\tesla}.}\label{fig3}
\end{figure*}

\textbf{Pulsed current measurements.} Since the continuous currents are expected to create significant Joule heating in these highly-resistive structures ($R \approx 1500\, \Omega$), we resort to pulsed measurements to address the influence of potential heating effects on the threshold current density. For this, the source-meter is hereafter replaced by a pulse generator whilst the rest of the setup is kept the same (see FIG. \ref{fig1} (c)). 200 \,\si{\nano\second} long square pulses are applied to excite the system, with a subsequent cooling time of 600 \,\si{\nano\second} without current. In FIG. \ref{fig3}, we present the results of the pulsed measurements at the same external field $\mu_{\mathrm{0}}H_{\mathrm{ext}}$ = 25 \,\si{\milli\tesla}. The time resolved $\mu$BLS intensity and frequency are presented in the color-coded plot of FIG. \ref{fig3} (a). The time interval ranging from 100 \,\si{\nano\second} to 300 \,\si{\nano\second} highlights the period where the pulsed excitation is turned on. Outside this interval, the pulse is turned off, and the measured $\mu$BLS intensity there is caused by thermal magnons, dark counts and possible ambient light. The pulsed excitation leads to an increase in the thermal magnons' intensity, which at the presented current density $J_{\mathrm{c}}$ = 1.87 $\times 10^{10}$ \,\si{\ampere\per\square\metre} is nearly twice the intensity of the thermal background. The data presented in FIG. \ref{fig3} (a) belongs to a measurement carried-out at a fixed $J_{\mathrm{c}}$. Sweeping $J_{\mathrm{c}}$ and thereafter averaging over 160 \,\si{\nano\second} of the pulse's total span of 200 \,\si{\nano\second} (the first and the last 20 \,\si{\nano\second} were excluded to ensure that the intensity is in the non-rising/non-decaying phase respectively), we construct the $\mu$BLS intensity map as function of BLS frequency and the applied current density (FIG. \ref{fig3} (b)), which can be directly compared to FIG. \ref{fig2} (a). In FIG. \ref{fig3} (b), the frequencies of the magnons detected are comparable to those from the continuous case, however the $J_{\mathrm{c}}$ values needed to achieve a significant increase above the thermal level are much lower in the pulsed case. In FIG. \ref{fig3} (c), we follow the same method of threshold current density extraction presented earlier in In FIG. \ref{fig2} (b). As shown from the extrapolation to zero inverse time-averaged $\mu$BLS intensities, the obtained threshold current density value from the pulsed measurement is then (0.292 $\pm$ 0.035) $\times 10^{10}$ \,\si{\ampere\per\square\metre}. This value is nearly six times less than the one extracted from the continuous current case, and represents a reduction of nearly two orders of magnitude comparison to most recent publications \cite{behera2024ultra,bainsla2025energy}. The significant change in threshold current density between continuous DC and pulsed measurements emphasizes that heat buildup can easily become prominent in such structures. The system of our study being micro-sized allows for higher lateral thermal resistance compared to a nano-sized spin Hall oscillator (considering consistency among other factors such as the substrate and the thickness). For a nanometer thick, micro-sized system, heat dissipation is mainly 1D into the the bulk of the substrate, whereas for a nano-sized system the lateral thermal conduction (e.g. through the thick electrical contact pads) is important, thus allowing for a cooler system. Temperature increase itself has been shown to increase the threshold current in spin Hall oscillators and similar bi-layers \cite{muralidhar2022optothermal}.

\begin{figure}[!ht]
\centering
\includegraphics[width=\columnwidth]{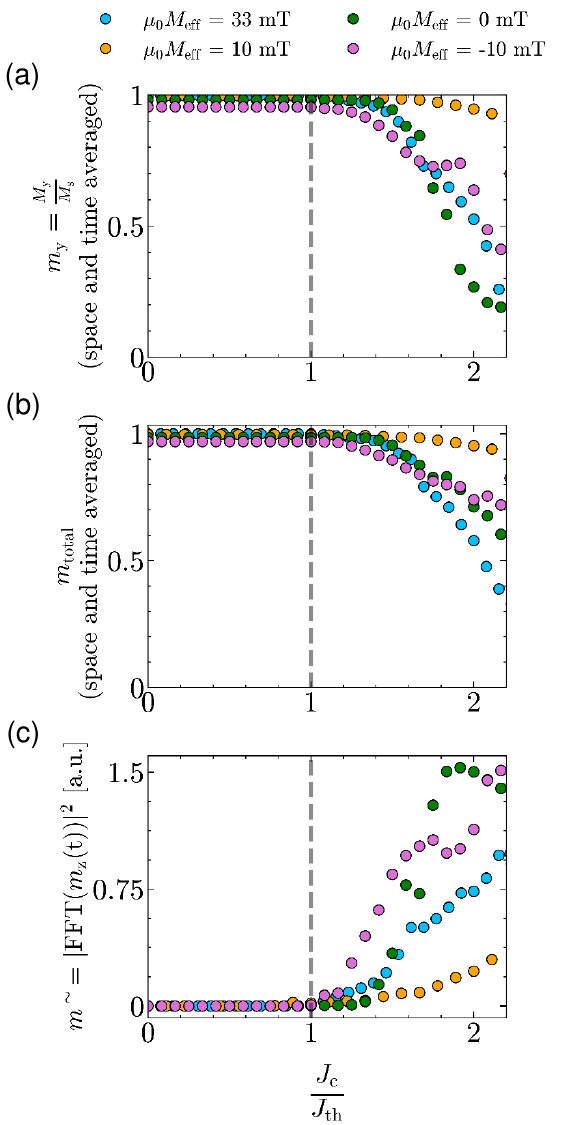}
\caption{Micromagnetic simulations with pulsed current: (a) Space and time averaged magnetization component $m_{\mathrm{y}}$ parallel to the external field, (b) Space and time averaged total magnetization $m_{\mathrm{total}}$ and (c) Normalized integrated FFT intensity $m^{\mathrm{\sim}}$ versus current density for $\mu_{\mathrm{0}}M_{\mathrm{eff}}$ = 33 \,\si{\milli\tesla}, $\mu_{\mathrm{0}}M_{\mathrm{eff}}$ = 10 \,\si{\milli\tesla}, $\mu_{\mathrm{0}}M_{\mathrm{eff}}$ = 0 \,\si{\milli\tesla} and $\mu_{\mathrm{0}}M_{\mathrm{eff}}$ = -10 \,\si{\milli\tesla}. The current densities are normalized separately to $J_{\mathrm{th}}$ for each $\mu_{\mathrm{0}}M_{\mathrm{eff}}$. Estimated for the respective shown cases of $\mu_{\mathrm{0}}M_{\mathrm{eff}}$, the threshold current densities are as follows: $J_{\mathrm{th}}$ = 1.3 $\times 10^{10}$ \,\si{\ampere\per\square\metre}, 0.9 $\times 10^{10}$ \,\si{\ampere\per\square\metre}, 1.2 $\times 10^{10}$ \,\si{\ampere\per\square\metre} and 1.2 $\times 10^{10}$ \,\si{\ampere\per\square\metre}. The external magnetic field across all simulations is fixed at $\mu_{\mathrm{0}}H_{\mathrm{ext}}$ = 25 \,\si{\milli\tesla}.}\label{fig4}
\end{figure}

\begin{figure*}[ht]
\centering
\includegraphics[width=\textwidth]{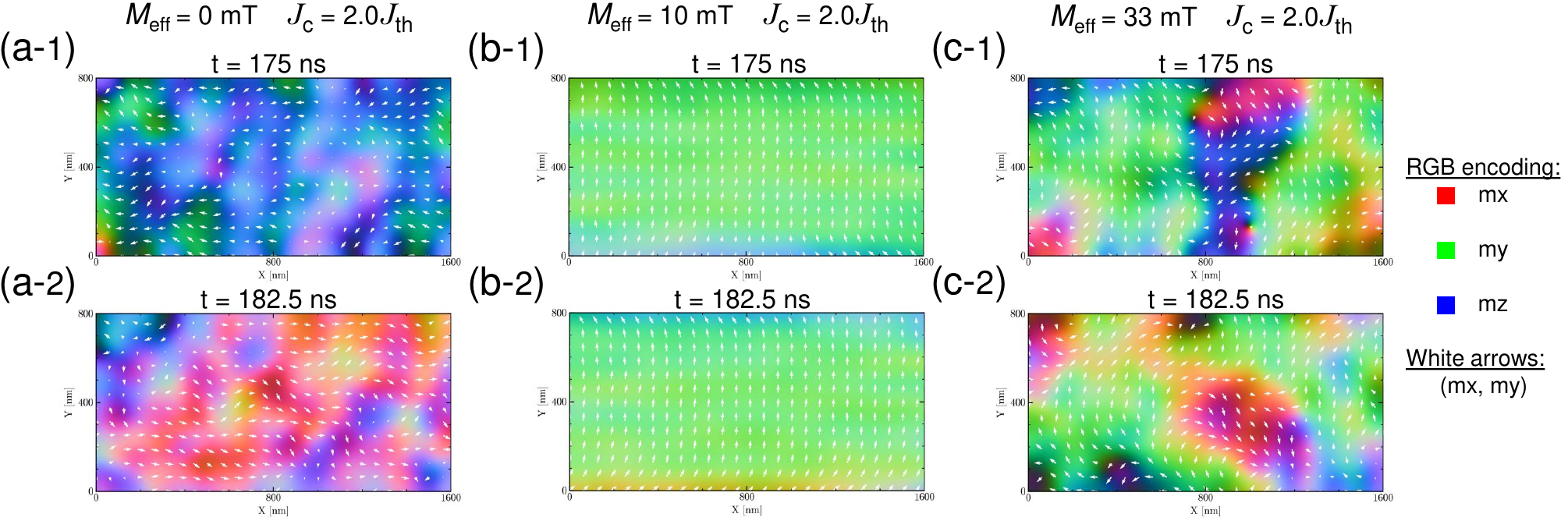}
\caption{Micromagnetic simulations with pulsed current: Spatial snapshots at t = 175 \,\si{\nano\second} and $J_{\mathrm{c}}$ = 2$J_{\mathrm{th}}$ for (a-1) $\mu_{\mathrm{0}}M_{\mathrm{eff}}$ = 0 \,\si{\milli\tesla} (b-1) $\mu_{\mathrm{0}}M_{\mathrm{eff}}$ = 10 \,\si{\milli\tesla} and (c-1) $\mu_{\mathrm{0}}M_{\mathrm{eff}}$ = 33 \,\si{\milli\tesla}. Spatial snapshots at t = 182.5 \,\si{\nano\second} and $J_{\mathrm{c}}$ = 2$J_{\mathrm{th}}$ for (a-2) $\mu_{\mathrm{0}}M_{\mathrm{eff}}$ = 0 \,\si{\milli\tesla} (b-2) $\mu_{\mathrm{0}}M_{\mathrm{eff}}$ = 10 \,\si{\milli\tesla} and (c-2) $\mu_{\mathrm{0}}M_{\mathrm{eff}}$ = 33 \,\si{\milli\tesla}. The RGB encoding and the white arrows used to represent the spatial evolution of the magnetization are shown to the right of the figure. The external magnetic field across all simulations is fixed at $\mu_{\mathrm{0}}H_{\mathrm{ext}}$ = 25 \,\si{\milli\tesla}.}\label{fig5}
\end{figure*}

\textbf{Micromagnetic simulations.} From a macrospin approach, as it was assumed in Eqn. \ref{eq:1}, zero $M_{\mathrm{eff}}$ should result in vanishing stabilized steady state magnetization oscillations above the threshold, since the current to destabilize the magnetization and the current to stabilize it against the magnetic field should be identical \cite{kurebayashi2026dynamical}. In experiment, we observe magnon intensities up to multiple times the value of the threshold current density itself, which negates the prediction made by the macrospin model. To investigate whether macrospin persists in describing magnetization dynamics in SHMOs, we resort to micromagnetic simulations using Mumax3 \cite{vansteenkiste2014design}. Material parameters, SHMO geometry and pulse length have values that agree with those from the experiments. For additional information regarding the setup of the simulations and the data analysis, refer to the \textbf{Micromagnetic simulations} subsection of the \textbf{Methods} section. In FIG. \ref{fig4} (a), the magnetization component $m_{\mathrm{y}}$ (space and time averaged in the 200 \,\si{\nano\second} pulse window, as displayed in the red shaded region in FIG. \ref{fig6} from the supplement, and normalized to $M_{\mathrm{s}}$) along the external magnetic field is presented versus the current density $J_{\mathrm{c}}$ (normalized to $J_{\mathrm{th}}$ for each $M_{\mathrm{eff}}$). The estimated values of $J_{\mathrm{th}}$ are shown in the caption of FIG. \ref{fig4}. Four different values of $M_{\mathrm{eff}}$ in the vicinity of zero are presented, namely $\mu_{\mathrm{0}}M_{\mathrm{eff}}$ = 33 \,\si{\milli\tesla}, $\mu_{\mathrm{0}}M_{\mathrm{eff}}$ = 10 \,\si{\milli\tesla}, $\mu_{\mathrm{0}}M_{\mathrm{eff}}$ = 0 \,\si{\milli\tesla} and $\mu_{\mathrm{0}}M_{\mathrm{eff}}$ = -10 \,\si{\milli\tesla}. Below $J_{\mathrm{c}}$ = $J_{\mathrm{th}}$, $m_{\mathrm{y}}$ remains at the maximum value of 1. As $J_{\mathrm{c}}$ becomes higher, that value starts to drop below 1 shifting towards 0 for all presented $M_{\mathrm{eff}}$ except for $\mu_{\mathrm{0}}M_{\mathrm{eff}}$ = 10 \,\si{\milli\tesla}. For $\mu_{\mathrm{0}}M_{\mathrm{eff}}$ = 10 \,\si{\milli\tesla}, the value stays close to 1 even above $J_{\mathrm{c}}$ = 2$J_{\mathrm{th}}$, which means that $m_{\mathrm{y}}$ is still mainly along the external field. This is not the case for $\mu_{\mathrm{0}}M_{\mathrm{eff}}$ = 33, 0 and -10 \,\si{\milli\tesla}, where $m_{\mathrm{y}}$ is becoming smaller, and is not mainly along the external field anymore. The magnetization along the external field ($m_{\mathrm{y}}$ in this case) is typically expected to remain close to the maximum value of 1 for current densities around the threshold, as that magnetization component is usually referred to as the static component. This fails to be the case here. As such, the space and time averaged total normalized magnetization $m_\mathrm{total}=\sqrt{(m_\mathrm{x}^2+m_\mathrm{y}^2+m_\mathrm{z}^2)}$ versus current density is considered next in FIG. \ref{fig4} (b). In the macrospin approximation, $m_\mathrm{total}$ is assumed to remain conserved at 1 for all situations. From FIG. \ref{fig4} (b), this is close to being the case for $\mu_{\mathrm{0}}M_{\mathrm{eff}}$ = 10 \,\si{\milli\tesla}, where $m_\mathrm{total}$ remains close to 1 for the studied range of current densities. However, $m_\mathrm{total}$ shifts far below 1 for $\mu_{\mathrm{0}}M_{\mathrm{eff}}$ = 33, 0 and -10 \,\si{\milli\tesla}. This means that the studied system at $J_\mathrm{c}>J_\mathrm{th}$ features complex dynamics involving finite wavelength spin waves (see FFT in FIG. \ref{fig6} in the supplement) or magnetic domains that cannot be described by a macrospin model (see FIG. \ref{fig5}). For $\mu_{\mathrm{0}}M_{\mathrm{eff}}$ = 10 \,\si{\milli\tesla}, the induced dynamics are still too weak to cause the macrospin breakdown. This is visible in FIG. \ref{fig4} (c), where we present a measure for the dynamics of the magnetization (normalized Fast Fourier Transform (FFT) intensity of $m_\mathrm{z}(t)$ integrated for frequencies starting from nearly 100 \,\si{\mega\hertz} up to 4 \,\si{\giga\hertz}, namely $m^{\mathrm{\sim}}$ = |FFT ($m_\mathrm{z}$(t))|$^2$). The temporal evolution of the dynamic magnetization component $m_\mathrm{z}$ and the corresponding FFT plots are included in FIG. \ref{fig6} in the \textbf{Micromagnetic simulations} subsection of the \textbf{Methods} section. As expected, $m^{\mathrm{\sim}}$ is null below the threshold, and starts to grow for higher current densities. For $\mu_{\mathrm{0}}M_{\mathrm{eff}}$ = 10 \,\si{\milli\tesla}, $m^{\mathrm{\sim}}$ only grows gradually with the current density. This is not the case for other values of $M_{\mathrm{eff}}$ where the growth of $m^{\mathrm{\sim}}$ is less regular and shows local minima. We can conclude that the behavior and the magnitude of the dynamics are completely different even for small changes of $M_{\mathrm{eff}}$, highlighting the huge influence of the latter on dynamic processes. These findings show that magnetization dynamics in a SHMO cannot be fully captured by a macrospin approximation. The breaking down of the macrospin approximation at current densities above the threshold $J_\mathrm{c}$ explains the presence of steady state magnetization oscillations even for absolute zero $M_{\mathrm{eff}}$. Worth noting, the assumptions made earlier for the threshold current density $J_{\mathrm{th}}$ in Eqn. \ref{eq:1} and Eqn. \ref{eq:2} still hold, as those equations assume very small magnetization oscillations, which is the case at very low current densities where macrospin still prevails.

The analysis above has demonstrated that finite wavelength spin waves and magnetic domains play a crucial role for the dynamics of SHMOs. In the FFT plots in FIG. \ref{fig6} (see \textbf{Micromagnetic simulations} subsection of the \textbf{Methods} section), finite wavelength spin waves are shown to be excited for all cases of $M_{\mathrm{eff}}$ at $J_{\mathrm{c}}$ = 2$J_{\mathrm{th}}$, though with varying intensities depending on $M_{\mathrm{eff}}$. These spin waves mostly span frequencies up to 2 \,\si{\giga\hertz}. More finite wavelength spin wave modes are clearly present for the the cases of $\mu_{\mathrm{0}}M_{\mathrm{eff}}$ = 0 \,\si{\milli\tesla} and $\mu_{\mathrm{0}}M_{\mathrm{eff}}$ = 33 \,\si{\milli\tesla}, which aligns with the results from FIG. \ref{fig4}, as these two cases exhibit the most deviation from macrospin. To illustrate this further and to address the formation of magnetic domains, we present spatial snapshots in FIG. \ref{fig5} for $\mu_{\mathrm{0}}M_{\mathrm{eff}}$ = 0 \,\si{\milli\tesla} (FIG. \ref{fig5} (a-1, a-2)), for $\mu_{\mathrm{0}}M_{\mathrm{eff}}$ = 10 \,\si{\milli\tesla} (FIG. \ref{fig5} (b-1, b-2)) and for $\mu_{\mathrm{0}}M_{\mathrm{eff}}$ = 33 \,\si{\milli\tesla} (FIG. \ref{fig5} (c-1, c-2)). The snapshots from the top row are taken at $t=$175 \,\si{\nano\second} into the simulations, and the ones from the bottom row 7.5 \,\si{\nano\second} later, at $t=$182.50 \,\si{\nano\second}. All snapshots were taken at a fixed current density of $J_{\mathrm{c}}$ = 2$J_{\mathrm{th}}$. The animations from which the snapshots were taken are supplemented as part of the \textbf{Micromagnetic simulations} subsection of the \textbf{Methods} section. In FIG. \ref{fig5} (a-1), it can be seen that the magnetization across the SHMO is not uniform and is broken into domains pointing in different directions, thus leading to a reduced total magnetization upon spatial averaging. Shifting $\mu_{\mathrm{0}}M_{\mathrm{eff}}$ to a slightly higher value of 10 \,\si{\milli\tesla} (see FIG. \ref{fig5} (b-1)) drastically changes this observed behavior. In FIG. \ref{fig5} (b-1), the magnetization appears to be more uniform throughout, retaining the macrospin behavior. Increasing $\mu_{\mathrm{0}}M_{\mathrm{eff}}$ to 33 \,\si{\milli\tesla} (see FIG. \ref{fig5} (c-1)), the domain distribution appears once again, but with a different profile in comparison to that from FIG. \ref{fig5} (a-1). In all three cases, the spatial distribution confirms the prior analysis from FIG. \ref{fig4} (a, b). For the studied current density range, macrospin breaks down for all simulated values of $M_{\mathrm{eff}}$, though the radicalness is different for each case. Comparing the magnetization at t = 175 \,\si{\nano\second} and t = 182.5 \,\si{\nano\second} for $\mu_{\mathrm{0}}M_{\mathrm{eff}}$ = 0 \,\si{\milli\tesla} and for $\mu_{\mathrm{0}}M_{\mathrm{eff}}$ = 33 \,\si{\milli\tesla} (FIG. \ref{fig5} (a-2) and (c-2), respectively), it can be clearly seen that the formed domains exhibit dynamicity with a strong change of the spatial magnetization distribution. This is shown to not be the case for $\mu_{\mathrm{0}}M_{\mathrm{eff}}$ = 10 \,\si{\milli\tesla}, where the change between the two snapshots (FIG. \ref{fig5} (b-1) and (b-2)) is nearly negligible in comparison to the other two cases of $M_{\mathrm{eff}}$. These results highlight the trend captured by $m^{\mathrm{\sim}}$ presented in FIG. \ref{fig4} (c), whereby the latter was shown to be stronger for $\mu_{\mathrm{0}}M_{\mathrm{eff}}$ = 0 \,\si{\milli\tesla} and $\mu_{\mathrm{0}}M_{\mathrm{eff}}$ = 33 \,\si{\milli\tesla}, than it was for $\mu_{\mathrm{0}}M_{\mathrm{eff}}$ = 10 \,\si{\milli\tesla}, thus highlighting the strong dynamicity of the formed domains floating all around space. The drastic changes in the dynamicity of the magnetic state in response to slight changes in $M_{\mathrm{eff}}$ highlights the importance of $M_{\mathrm{eff}}$ tuning and engineering in SHMOs, as that allows for both low threshold dynamics and control over the magnetic state of the system. This enables possibilities to use SHMOs to study dynamic stability \cite{kurebayashi2026dynamical,karadza2026dynamical} and/or antimagnon physics \cite{harms2022enhanced,harms2024antimagnonics,karadza2026dynamical,Wang2026direct}.  

In conclusion, we have demonstrated via $\mu$BLS measurements that near zero effective magnetization can drastically reduce the needed current densities for driving magnetization dynamics in spin Hall systems. Energy efficiency being a major point to address in such systems, we highlighted the difference in threshold current densities for continuous current measurements versus pulsed ones. The threshold current density for pulsed excitations is noticeably lower, with it exhibiting a reduction of nearly two orders of magnitude in comparison to earlier reported values. Micromagnetic simulations thereafter show that dynamics in SHMOs can go far beyond the macrospin approximation at higher currents, and highlight the effect of engineering $M_{\mathrm{eff}}$ also on the magnetization dynamics beyond macrospin. Our findings present SHMOs as a promising platform for energy efficient magnetization control, and candidate these devices for studying dynamic stability and antimagnon physics.

\section*{Methods}

\vspace{0.6cm}

\subsection*{Stack Fabrication}

\subsubsection{Thin film preparation}
The thin film was grown on a thermally-oxidized Silicon (Si) substrate by employing magnetron sputtering with the base pressure below 2 $\times 10^{-7}$ \,\si{\pascal}. The stack is constituted as such: Substrate|W (4 \,\si{\nano\metre})|CoFeB (1.7 \,\si{\nano\metre})|MgO (3 \,\si{\nano\metre})|Ta (1.5 \,\si{\nano\metre}). All layers were deposited at ambient temperature under the Argon (Ar) gas pressure of approximately 5 \,\si{\mTorr}. The W, CoFeB (atomic composition of Co$_\mathrm{20}$Fe$_\mathrm{60}$B$_\mathrm{20}$) and Ta targets were sputtered using DC power supplies, while the MgO target was sputtered using an rf power supply. The thin film was annealed at 300 \,\si{\degreeCelsius} for 1 hour in a vacuum furnace with the base pressure below 1 $\times 10^{-4}$ \,\si{\pascal}.

\subsubsection{Microfabrication process}

The thin films were patterned into a rectangle with 10 \,\si{\micro\metre} width and 40 \,\si{\micro\metre} length (signal line) by standard optical lithography and Ar ion milling. Contact pads of Cr (20 \,\si{\nano\metre})|Au (100 \,\si{\nano\metre}) were formed using standard optical lithography, ion beam sputter-deposition of Cr and Au and lift-off techniques. The signal lines were made narrow by electron beam lithography and Ar ion milling, resulting in the rectangular-shaped bar with 1 \,\si{\micro\metre} width and 1.8 \,\si{\micro\metre} length Si-O capping layer was deposited to prevent from the degradation of narrow signal lines. Extended parts of coplanar waveguides, made of Ta (2 \,\si{\nano\metre})|Au(100 \,\si{\nano\metre}), were fabricated using standard optical lithography, sputter-deposition of Ta and Au and lift-off techniques.

\subsection*{Micromagnetic simulations}

\begin{figure}[!ht]
\centering
\includegraphics[width=\columnwidth]{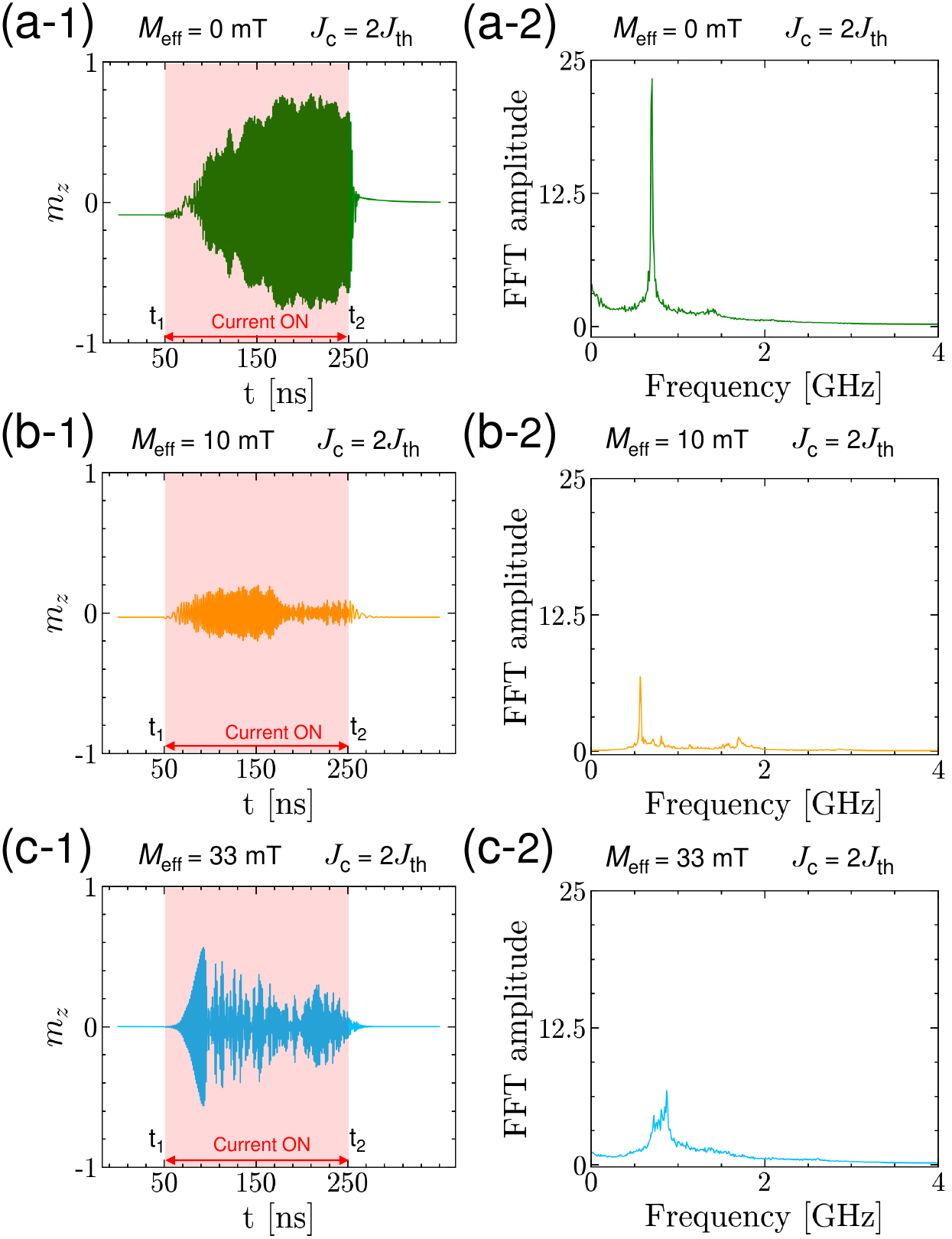}
\caption{Micromagnetic simulations with pulsed current: Space averaged temporal dynamic magnetization component $m_{\mathrm{z}}$ (normalized to $M_{\mathrm{s}}$) at $J_{\mathrm{c}}$ = 2$J_{\mathrm{th}}$ for (a-1) $\mu_{\mathrm{0}}M_{\mathrm{eff}}$ = 0 \,\si{\milli\tesla}, (b-1) $\mu_{\mathrm{0}}M_{\mathrm{eff}}$ = 10 \,\si{\milli\tesla} and (c-1) $\mu_{\mathrm{0}}M_{\mathrm{eff}}$ = 33 \,\si{\milli\tesla}. The FFT corresponding to each case of respective $M_{\mathrm{eff}}$ is presented in the adjacent column featuring (a-2, b-2, c-2). The external magnetic field across all simulations is fixed at $\mu_{\mathrm{0}}H_{\mathrm{ext}}$ = 25 \,\si{\milli\tesla}.}\label{fig6}
\end{figure}

Micromagnetic simulations are performed by means of Mumax3 \cite{vansteenkiste2014design} deployed onto the platform Aithericon \cite{aithericon}. Magnetic parameters, device dimensions and pulse definition are taken from the experiments and material characterization. As such, a saturation magnetization $M_{\mathrm{s}}$ = 521$\times10^3$ \,\si{\ampere\per\metre}, exchange stiffness $A_{ex}$ = 15$\times10^{-12}$ \,\si{\joule\per\metre} (typical value for CoFeB \cite{devolder2016exchange,choi2020exchange}), a damping constant $\alpha$ = 0.045 and a uniaxial out-of-plane anisotropy constant $K_\mathrm{u1}$ $\in$ [161950, 167944, 170551.3, 173156.7] \,\si{\joule\per\cubic\metre} resulting in $\mu_{\mathrm{0}}M_{\mathrm{eff}}$ $\in$ [33, 10, 0, -10] \,\si{\milli\tesla}, respectively, were chosen to model the CoFeB layer. The CoFeB layer's dimensions used in the simulations are 1600$\times$800$\times$1.7 \,\si{\nano\meter\cubed}. The lateral dimensions deviate slightly from the real structure's dimensions to allow the cell sizes in Mumax3 to align and facilitate the simulations. The thickness, however, is exactly matched to the real system's thickness because it directly affects the threshold current density. The magnetic field is applied along the short axis with a magnitude $\mu_{\mathrm{0}}H_{\mathrm{ext}}$ = 25 \,\si{\milli\tesla}. The simulations were performed in the absence of thermal effects ($T$ = 0 \,\si{\kelvin}) to reduce simulation time. To excite magnetization dynamics, 200 \,\si{\nano\second} long pulses are thereafter applied to the system, where a Heaviside step-function was chosen. The simulations are allowed long enough elapsed time (350 \,\si{\nano\second} run time) to visualize magnetization dynamics prior to, during and post pulse application. The temporal averaging of the magnetization component $m_{\mathrm{y}}$ as well as the total magnetization $m_{\mathrm{total}}$ (FIG. \ref{fig4} (a, b)) was computed during the 200 \,\si{\nano\second} pulse window only (see red shaded region in FIG. \ref{fig6} for (a-1, b-1, c-1)). Outside the pulse (at zero currents), the dynamics are dead. On the left column of FIG. \ref{fig6}, the temporal evolution of the dynamic magnetization component $m_{\mathrm{z}}$ (space and time averaged and normalized to $M_{\mathrm{s}}$) is presented at $J_{\mathrm{c}}$ = 2$J_{\mathrm{th}}$ for (a-1) $\mu_{\mathrm{0}}M_{\mathrm{eff}}$ = 0 \,\si{\milli\tesla}, (b-1) $\mu_{\mathrm{0}}M_{\mathrm{eff}}$ = 10 \,\si{\milli\tesla} and (c-1) $\mu_{\mathrm{0}}M_{\mathrm{eff}}$ = 33 \,\si{\milli\tesla}. The threshold current densities in the simulations for each case of $M_{\mathrm{eff}}$ are estimated based on the definition of the threshold current density itself \cite{slavin2009nonlinear}, that being the current density at which the system's intrinsic damping is fully compensated, and the point beyond which the magnetization oscillations start to grow instead of decay in time. Those current densities are as follows: $J_{\mathrm{th}}$ = 1.3 $\times 10^{10}$ \,\si{\ampere\per\square\metre} for $\mu_{\mathrm{0}}M_{\mathrm{eff}}$ = 33 \,\si{\milli\tesla}, 0.9 $\times 10^{10}$ \,\si{\ampere\per\square\metre} for $\mu_{\mathrm{0}}M_{\mathrm{eff}}$ = 10 \,\si{\milli\tesla} and 1.2 $\times 10^{10}$ \,\si{\ampere\per\square\metre} for $\mu_{\mathrm{0}}M_{\mathrm{eff}}$ = 0 \,\si{\milli\tesla}. At $J_{\mathrm{c}}$ = 2$J_{\mathrm{th}}$, $m_{\mathrm{z}}$ exhibits different oscillatory behavior and different amplitudes depending on $M_{\mathrm{eff}}$ (see the difference between the three plots in the left column). The FFT of the magnetization component $m_{\mathrm{z}}$ (right column of FIG. \ref{fig6}) is computed in every cell individually, then averaged over the full space, rather than computed directly for the averaged magnetization, to capture the correct collective magnetization behavior (including shorter wavelength spin waves), instead of getting smeared out data due to averaging. This is presented in FIG. \ref{fig6} (a-2, b-2, c-2) for the same previous conditions of $J_{\mathrm{c}}$ and $M_{\mathrm{eff}}$. Multiple modes are excited in each of the cases, with modes above the fundamental mode belonging to finite wavevector spin-wave modes. The mode appearing near zero frequency for $\mu_{\mathrm{0}}M_{\mathrm{eff}}$ = 0 \,\si{\milli\tesla} is due to the formation of some static spatial domains.
The animations, from which the snapshots of FIG. \ref{fig5} were taken, are attached with the manuscript as part of the supplement. Those animations show the temporal magnetization evolution in space between $t=$175 \,\si{\nano\second} and $t=$185 \,\si{\nano\second} for $\mu_{\mathrm{0}}M_{\mathrm{eff}}$ = 0 \,\si{\milli\tesla} (animation1.mp4), $\mu_{\mathrm{0}}M_{\mathrm{eff}}$ = 10 \,\si{\milli\tesla} (animation2.mp4) and $\mu_{\mathrm{0}}M_{\mathrm{eff}}$ = 33 \,\si{\milli\tesla} (animation3.mp4).

\section*{Data availability}
All data that supports the plots within this paper and other findings of this study are available from the corresponding authors upon reasonable request.

\section*{Code availability}
The codes that generate the plots within this paper and other findings of this study are available from the corresponding authors upon reasonable request.

\section*{References}
\bibliography{ref}

@article{fischer2017experimental,
  title={Experimental prototype of a spin-wave majority gate},
  author={Fischer, Thomas and Kewenig, M and Bozhko, DA and Serga, AA and Syvorotka, II and Ciubotaru, Florin and Adelmann, Christoph and Hillebrands, B and Chumak, Andrii V},
  journal={Applied Physics Letters},
  volume={110},
  number={15},
  year={2017},
  publisher={AIP Publishing}
}

@article{csaba2017perspectives,
  title={Perspectives of using spin waves for computing and signal processing},
  author={Csaba, Gy{\"o}rgy and Papp, {\'A}d{\'a}m and Porod, Wolfgang},
  journal={Physics Letters A},
  volume={381},
  number={17},
  pages={1471--1476},
  year={2017},
  publisher={Elsevier}
}

@article{mahmoud2023two,
  title={Two cascaded spin wave majority gates operation under continuous and pulse modes},
  author={Mahmoud, Abdulqader Nael and Ciubotaru, Florin and Vanderveken, Frederic and Adelmann, Christoph and Cotofana, Sorin and Hamdioui, Said},
  journal={IEEE Transactions on Circuits and Systems II: Express Briefs},
  volume={71},
  number={4},
  pages={1919--1923},
  year={2023},
  publisher={IEEE}
}

@article{chumak2022advances,
  title={Advances in magnetics roadmap on spin-wave computing},
  author={Chumak, Andrii V and Kabos, Pavel and Wu, Mingzhong and Abert, Claas and Adelmann, Christoph and Adeyeye, Adekunle Olusola and {\AA}kerman, Johan and Aliev, Farkhad G and Anane, Abdelmadjid and Awad, Ahmad and others},
  journal={IEEE Transactions on Magnetics},
  volume={58},
  number={6},
  pages={1--72},
  year={2022},
  publisher={IEEE}
}

@article{levchenko2024review,
  title={Review on spin-wave RF applications},
  author={Levchenko, Khrystyna O and Dav{\'\i}dkov{\'a}, Krist{\`y}na and Mikkelsen, Jan and Chumak, Andrii V},
  journal={arXiv preprint arXiv:2411.19212},
  year={2024}
}

@article{kohl2025identification,
  title={Identification and minimization of losses in microscaled spin-wave transducers},
  author={Kohl, Felix and Heinz, Bj{\"o}rn and Papp, {\'A}d{\'a}m and Csaba, Gy{\"o}rgi and Pirro, Philipp},
  journal={arXiv preprint arXiv:2505.08656},
  year={2025}
}

@article{demidov2012magnetic,
  title={Magnetic nano-oscillator driven by pure spin current},
  author={Demidov, Vladislav E and Urazhdin, Sergei and Ulrichs, Henning and Tiberkevich, Vasyl and Slavin, Andrei and Baither, Dietmar and Schmitz, Guido and Demokritov, Sergej O},
  journal={Nature materials},
  volume={11},
  number={12},
  pages={1028--1031},
  year={2012},
  publisher={Nature Publishing Group UK London}
}

@article{awad2017long,
  title={Long-range mutual synchronization of spin Hall nano-oscillators},
  author={Awad, AA and D{\"u}rrenfeld, Ph and Houshang, A and Dvornik, M and Iacocca, Ezio and Dumas, RK and {\AA}kerman, Johan},
  journal={Nature Physics},
  volume={13},
  number={3},
  pages={292--299},
  year={2017},
  publisher={Nature Publishing Group UK London}
}

@article{chen2016spin,
  title={Spin-torque and spin-Hall nano-oscillators},
  author={Chen, Tingsu and Dumas, Randy K and Eklund, Anders and Muduli, Pranaba K and Houshang, Afshin and Awad, Ahmad A and D{\"u}rrenfeld, Philipp and Malm, B Gunnar and Rusu, Ana and {\AA}kerman, Johan},
  journal={Proceedings of the IEEE},
  volume={104},
  number={10},
  pages={1919--1945},
  year={2016},
  publisher={IEEE}
}

@article{zahedinejad2020two,
  title={Two-dimensional mutually synchronized spin Hall nano-oscillator arrays for neuromorphic computing},
  author={Zahedinejad, Mohammad and Awad, Ahmad A and Muralidhar, Shreyas and Khymyn, Roman and Fulara, Himanshu and Mazraati, Hamid and Dvornik, Mykola and {\AA}kerman, Johan},
  journal={Nature nanotechnology},
  volume={15},
  number={1},
  pages={47--52},
  year={2020},
  publisher={Nature Publishing Group UK London}
}

@article{zahedinejad2022memristive,
  title={Memristive control of mutual spin Hall nano-oscillator synchronization for neuromorphic computing},
  author={Zahedinejad, Mohammad and Fulara, Himanshu and Khymyn, Roman and Houshang, Afshin and Dvornik, Mykola and Fukami, Shunsuke and Kanai, Shun and Ohno, Hideo and {\AA}kerman, Johan},
  journal={Nature materials},
  volume={21},
  number={1},
  pages={81--87},
  year={2022},
  publisher={Nature Publishing Group UK London}
}

@article{haidar2019single,
  title={A single layer spin-orbit torque nano-oscillator},
  author={Haidar, Mohammad and Awad, Ahmad A and Dvornik, Mykola and Khymyn, Roman and Houshang, Afshin and {\AA}kerman, Johan},
  journal={Nature communications},
  volume={10},
  number={1},
  pages={2362},
  year={2019},
  publisher={Nature Publishing Group UK London}
}

@article{bainsla2025energy,
  title={Energy-Efficient Single Layer Spin Hall Nano-Oscillators Driven by Berry Curvature},
  author={Bainsla, Lakhan and Sakuraba, Yuya and Kumar, Akash and Chaurasiya, Avinash Kumar and Masuda, Keisuke and Suwannaharn, Nattamon and Awad, Ahmad A and Behera, Nilamani and Khymyn, Roman and Sasaki, Taisuke and others},
  journal={ACS nano},
  volume={19},
  number={19},
  pages={18534--18544},
  year={2025},
  publisher={ACS Publications}
}

@article{behera2024ultra,
  title={Ultra-Low Current 10 nm Spin Hall Nano-Oscillators},
  author={Behera, Nilamani and Chaurasiya, Avinash Kumar and Gonz{\'a}lez, Victor H and Litvinenko, Artem and Bainsla, Lakhan and Kumar, Akash and Khymyn, Roman and Awad, Ahmad A and Fulara, Himanshu and {\AA}kerman, Johan},
  journal={Advanced Materials},
  volume={36},
  number={5},
  pages={2305002},
  year={2024},
  publisher={Wiley Online Library}
}

@article{jungwirth2012spin,
  title={Spin Hall effect devices},
  author={Jungwirth, Tomas and Wunderlich, J{\"o}rg and Olejn{\'\i}k, Kamil},
  journal={Nature materials},
  volume={11},
  number={5},
  pages={382--390},
  year={2012},
  publisher={Nature Publishing Group UK London}
}

@article{hao2015beta,
  title={Beta ($\beta$) tungsten thin films: Structure, electron transport, and giant spin Hall effect},
  author={Hao, Qiang and Chen, Wenzhe and Xiao, Gang},
  journal={Applied Physics Letters},
  volume={106},
  number={18},
  year={2015},
  publisher={AIP Publishing}
}

@article{behera2022energy,
  title={Energy-Efficient W 100- x Ta x/Co-Fe-B/MgO Spin Hall Nano-Oscillators},
  author={Behera, Nilamani and Fulara, Himanshu and Bainsla, Lakhan and Kumar, Akash and Zahedinejad, Mohammad and Houshang, Afshin and {\AA}kerman, Johan},
  journal={Physical Review Applied},
  volume={18},
  number={2},
  pages={024017},
  year={2022},
  publisher={APS}
}

@article{slavin2009nonlinear,
  title={Nonlinear auto-oscillator theory of microwave generation by spin-polarized current},
  author={Slavin, Andrei and Tiberkevich, Vasil},
  journal={IEEE Transactions on Magnetics},
  volume={45},
  number={4},
  pages={1875--1918},
  year={2009},
  publisher={IEEE}
}

@article{demidov2014synchronization,
  title={Synchronization of spin Hall nano-oscillators to external microwave signals},
  author={Demidov, VE and Ulrichs, H and Gurevich, SV and Demokritov, SO and Tiberkevich, VS and Slavin, AN and Zholud, A and Urazhdin, S},
  journal={Nature communications},
  volume={5},
  number={1},
  pages={3179},
  year={2014},
  publisher={Nature Publishing Group UK London}
}

@article{li2020recent,
  title={Recent progress on excitation and manipulation of spin-waves in spin Hall nano-oscillators},
  author={Li, Liyuan and Chen, Lina and Liu, Ronghua and Du, Youwei},
  journal={Chinese Physics B},
  volume={29},
  number={11},
  pages={117102},
  year={2020},
  publisher={IOP Publishing}
}

@article{choi2022voltage,
  title={Voltage-driven gigahertz frequency tuning of spin Hall nano-oscillators},
  author={Choi, Jong-Guk and Park, Jaehyeon and Kang, Min-Gu and Kim, Doyoon and Rieh, Jae-Sung and Lee, Kyung-Jin and Kim, Kab-Jin and Park, Byong-Guk},
  journal={Nature communications},
  volume={13},
  number={1},
  pages={3783},
  year={2022},
  publisher={Nature Publishing Group UK London}
}

@article{yasuda2017current,
  title={Current-nonlinear Hall effect and spin-orbit torque magnetization switching in a magnetic topological insulator},
  author={Yasuda, K and Tsukazaki, A and Yoshimi, R and Kondou, K and Takahashi, KS and Otani, Y and Kawasaki, M and Tokura, Y},
  journal={Physical review letters},
  volume={119},
  number={13},
  pages={137204},
  year={2017},
  publisher={APS}
}

@article{liu2013spectral,
  title={Spectral characteristics of the microwave emission by the spin Hall nano-oscillator},
  author={Liu, RH and Lim, WL and Urazhdin, S},
  journal={Physical review letters},
  volume={110},
  number={14},
  pages={147601},
  year={2013},
  publisher={APS}
}

@article{haidar2021compositional,
  title={Compositional effect on auto-oscillation behavior of Ni100- xFex/Pt spin Hall nano-oscillators},
  author={Haidar, Mohammad and Mazraati, Hamid and D{\"u}rrenfeld, Philipp and Fulara, Himanshu and Ranjbar, Mojtaba and {\AA}kerman, J},
  journal={Applied Physics Letters},
  volume={118},
  number={1},
  year={2021},
  publisher={AIP Publishing}
}

@article{demidov2011control,
  title={Control of magnetic fluctuations by spin current},
  author={Demidov, Vladislav E and Urazhdin, Sergei and Edwards, ERJ and Stiles, Mark D and McMichael, Robert D and Demokritov, Sergej O},
  journal={Physical review letters},
  volume={107},
  number={10},
  pages={107204},
  year={2011},
  publisher={APS}
}

@article{taniguchi2016instability,
  title={Instability analysis of spin-torque oscillator with an in-plane magnetized free layer and a perpendicularly magnetized pinned layer},
  author={Taniguchi, Tomohiro and Kubota, Hitoshi},
  journal={Physical Review B},
  volume={93},
  number={17},
  pages={174401},
  year={2016},
  publisher={APS}
}

@article{zhu2020threshold,
  title={Threshold current density for perpendicular magnetization switching through spin-orbit torque},
  author={Zhu, Daoqian and Zhao, Weisheng},
  journal={Physical Review Applied},
  volume={13},
  number={4},
  pages={044078},
  year={2020},
  publisher={APS}
}

@article{liu2012spin,
  title={Spin-torque switching with the giant spin Hall effect of tantalum},
  author={Liu, Luqiao and Pai, Chi-Feng and Li, Y and Tseng, HW and Ralph, DC and Buhrman, RA},
  journal={Science},
  volume={336},
  number={6081},
  pages={555--558},
  year={2012},
  publisher={American Association for the Advancement of Science}
}

@article{haidar2016controlling,
  title={Controlling Gilbert damping in a YIG film using nonlocal spin currents},
  author={Haidar, M and D{\"u}rrenfeld, P and Ranjbar, M and Balinsky, M and Fazlali, M and Dvornik, M and Dumas, RK and Khartsev, Sergiy and {\AA}kerman, Johan},
  journal={Physical Review B},
  volume={94},
  number={18},
  pages={180409},
  year={2016},
  publisher={APS}
}

@article{collet2016generation,
  title={Generation of coherent spin-wave modes in yttrium iron garnet microdiscs by spin--orbit torque},
  author={Collet, Martin and De Milly, Xavier and d’Allivy Kelly, Olivier and Naletov, Vladimir V and Bernard, Rozenn and Bortolotti, Paolo and Ben Youssef, J and Demidov, VE and Demokritov, SO and Prieto, Jose Luis and others},
  journal={Nature communications},
  volume={7},
  number={1},
  pages={10377},
  year={2016},
  publisher={Nature Publishing Group UK London}
}

@article{fulara2020giant,
  title={Giant voltage-controlled modulation of spin Hall nano-oscillator damping},
  author={Fulara, Himanshu and Zahedinejad, Mohammad and Khymyn, Roman and Dvornik, Mykola and Fukami, Shunsuke and Kanai, Shun and Ohno, Hideo and {\AA}kerman, Johan},
  journal={Nature communications},
  volume={11},
  number={1},
  pages={4006},
  year={2020},
  publisher={Nature Publishing Group UK London}
}

@article{shi2018fast,
  title={Fast low-current spin-orbit-torque switching of magnetic tunnel junctions through atomic modifications of the free-layer interfaces},
  author={Shi, Shengjie and Ou, Yongxi and Aradhya, SV and Ralph, DC and Buhrman, RA},
  journal={Physical Review Applied},
  volume={9},
  number={1},
  pages={011002},
  year={2018},
  publisher={APS}
}

@article{vansteenkiste2014design,
  title={The design and verification of MuMax3},
  author={Vansteenkiste, Arne and Leliaert, Jonathan and Dvornik, Mykola and Helsen, Mathias and Garcia-Sanchez, Felipe and Van Waeyenberge, Bartel},
  journal={AIP advances},
  volume={4},
  number={10},
  year={2014},
  publisher={AIP Publishing}
}

@article{muralidhar2022optothermal,
  title={Optothermal control of spin Hall nano-oscillators},
  author={Muralidhar, Shreyas and Houshang, Afshin and Alem{\'a}n, Ademir and Khymyn, Roman and Awad, Ahmad A and {\AA}kerman, Johan},
  journal={Applied Physics Letters},
  volume={120},
  number={26},
  year={2022},
  publisher={AIP Publishing}
}

@article{guckelhorn2021magnon,
  title={Magnon transport in Y 3 Fe 5 O 12/Pt nanostructures with reduced effective magnetization},
  author={G{\"u}ckelhorn, Janine and Wimmer, Tobias and M{\"u}ller, Manuel and Gepr{\"a}gs, Stephan and H{\"u}bl, Hans and Gross, Rudolf and Althammer, Matthias},
  journal={Physical Review B},
  volume={104},
  number={18},
  pages={L180410},
  year={2021},
  publisher={APS}
}

@article{divinskiy2019controlled,
  title={Controlled nonlinear magnetic damping in spin-Hall nano-devices},
  author={Divinskiy, Boris and Urazhdin, Sergei and Demokritov, Sergej O and Demidov, Vladislav E},
  journal={Nature communications},
  volume={10},
  number={1},
  pages={5211},
  year={2019},
  publisher={Nature Publishing Group UK London}
}

@article{birt2013brillouin,
  title={Brillouin light scattering spectra as local temperature sensors for thermal magnons and acoustic phonons},
  author={Birt, Daniel R and An, Kyongmo and Weathers, Annie and Shi, Li and Tsoi, Maxim and Li, Xiaoqin},
  journal={Applied Physics Letters},
  volume={102},
  number={8},
  year={2013},
  publisher={AIP Publishing}
}

@article{harms2022enhanced,
  title={Enhanced magnon spin current using the bosonic Klein paradox},
  author={Harms, JS and Yuan, HY and Duine, Rembert A},
  journal={Physical Review Applied},
  volume={18},
  number={6},
  pages={064026},
  year={2022},
  publisher={APS}
}

@article{harms2024antimagnonics,
  title={Antimagnonics},
  author={Harms, JS and Yuan, HY and Duine, Rembert A},
  journal={AIP Advances},
  volume={14},
  number={2},
  year={2024},
  publisher={AIP Publishing}
}

@article{yuan2023magnonic,
  title={Magnonic Klein and acausal tunneling enabled by breaking the anti parity-time symmetry in antiferromagnets},
  author={Yuan, Shaohua and Sui, Chaowei and Fan, Zhengduo and Berakdar, Jamal and Xue, Desheng and Jia, Chenglong},
  journal={Communications Physics},
  volume={6},
  number={1},
  pages={95},
  year={2023},
  publisher={Nature Publishing Group UK London}
}

@article{karadza2026dynamical,
  title={Dynamical Stabilization of Inverted Magnetization and Antimagnons by Spin Injection in an Extended Magnetic System},
  author={Karadza, Emir and Wang, Hanchen and Kercher, Niklas and Noel, Paul and Legrand, William and Schlitz, Richard and Gambardella, Pietro},
  journal={arXiv preprint arXiv:2601.09569},
  year={2026}
}

@article{Wang2026direct,
  title={Direct Observation of Antimagnons with Inverted Dispersion},
  author={Wang, Hanchen and Hu, Junfeng and Song, Wenjie and Bassant, Artim L. and Wang, Jinlong and Peng, Haishen and Karadza, Emir and Noel, Paul and Legrand, William and Schlitz, Richard and Chen, Jilei and Liu, Song and Yu, Dapeng and Ansermet, Jean-Philippe and Duine, Rembert A. and Gambardella, Pietro and Yu, Haiming},
  journal={arXiv preprint arXiv:2601.15231},
  year={2026}
}

@article{edwards2012parametric,
  title={Parametric excitation of magnetization oscillations controlled by pure spin current},
  author={Edwards, ERJ and Ulrichs, H and Demidov, VE and Demokritov, SO and Urazhdin, S},
  journal={Physical Review B—Condensed Matter and Materials Physics},
  volume={86},
  number={13},
  pages={134420},
  year={2012},
  publisher={APS}
}

@article{aithericon,
  title={},
  author={},
  journal={},
  volume={},
  number={},
  pages={},
  year={See www.aithericon.com for further information and details},
  publisher={}
}

@article{kurebayashi2026dynamical,
  title={Dynamical stability by spin transfer in nearly isotropic magnets},
  author={Kurebayashi, Hidekazu and Barker, Joseph and Yamazaki, Takumi and Kushwaha, Varun K and Stenning, Kilian D and Youel, Harry and Hou, Xueyao and Dion, Troy and Prestwood, Daniel and Bauer, Gerrit EW and others},
  journal={Nature Materials},
  pages={1--8},
  year={2026},
  publisher={Nature Publishing Group UK London}
}

@article{choi2020exchange,
  title={Exchange stiffness and damping constants of spin waves in CoFeB films},
  author={Choi, Gyung-Min},
  journal={Journal of Magnetism and Magnetic Materials},
  volume={516},
  pages={167335},
  year={2020},
  publisher={Elsevier}
}

@article{devolder2016exchange,
  title={Exchange stiffness in ultrathin perpendicularly magnetized CoFeB layers determined using the spectroscopy of electrically excited spin waves},
  author={Devolder, T and Kim, J-V and Nistor, L and Sousa, R and Rodmacq, B and Di{\'e}ny, B},
  journal={Journal of Applied Physics},
  volume={120},
  number={18},
  year={2016},
  publisher={AIP Publishing}
}

\section*{Acknowledgements}
This work has been  supported  by the European Research Council within the Starting Grant No. 101042439 "CoSpiN" and the Deutsche Forschungsgemeinschaft (DFG, German Research Foundation) - TRR 173 - 268565370" (project B01).\\
The device fabrication was partly carried out at the Cooperative Research and Development Center for Advanced Materials, IMR, Tohoku University, Japan. Financial support was provided by JSPS KAKENHI Grant-in-Aid for Scientific Research (A) (JP23H00232), and MEXT Initiative to Establish Next-generation Novel Integrated Circuits Centers (X-NICS) Grant Number JPJ011438.

\section*{Author contributions}
A.K. underwent all measurements, analyzed all experimental data and performed all micromagnetic simulations under the supervision and guidance of P.P. and A.H. B.H. supported A.K. in the measurements. T.S., V.K and X.H. carried out the film deposition and device fabrication. H.K. and T.S. supplied the chip/sample onto which the study was carried out. K.Y. assisted with insightful suggestions and explanations from a theoretical perspective. A.K., H.K., K.Y., B.H., A.H., T.S. and P.P. discussed the results and commented on the manuscript.

\section*{Competing interests}
The authors declare no competing interests.

\end{document}